\documentclass[12pt,leqno]{article} 

\usepackage[utf8]{inputenc} 

\usepackage{hyperref}
\hypersetup{
    unicode=false,          
    pdftitle={Spring Transformation},    
    pdfkeywords={curvature}, 
    colorlinks=true,       
    pdfborder=0 0 0,
    linkcolor=blue,          
    citecolor=blue,        
    filecolor=blue,      
    urlcolor=blue,           
}

\usepackage{amsthm}
\usepackage{amsfonts}
\usepackage{amsmath}
\usepackage{mathtools}

\usepackage{algorithm}
\usepackage[noend]{algpseudocode}

 \theoremstyle{definition}

 \theoremstyle{remark}

 \numberwithin{equation}{section}
\usepackage{geometry} 
\usepackage{graphicx} 
\usepackage{placeins}
\usepackage{xcolor}

\usepackage{booktabs} 
\usepackage{array} 
\usepackage{paralist} 
\usepackage{verbatim} 
\usepackage{subfig} 
\usepackage[font=small,labelfont=small]{caption}

\usepackage{fancyhdr} 
\usepackage[nottoc,notlof,notlot]{tocbibind} 
\usepackage[titles,subfigure]{tocloft} 
\usepackage{todonotes}

\newcommand{\cT}{{\cal{T}}}

\newcommand{\R}{\mathbf{R}}

\newcommand{\N}{\mathbf{N}}

\def\co{\colon\thinspace}

\makeatletter
\def\BState{\State\hskip-\ALG@thistlm}
\makeatother
\pdfoutput=1

\title{Smoothing Game}
\author{Dimitris Vartziotis \and Doris Bohnet \and Benjamin Himpel}
\begin{document}
\maketitle

\section{Introduction}

In order to improve meshes for finite element simulations they need to be smoothed by relocating the vertices. This goal is usually simplified by trying to make each element as regular as possible. There are different methods for achieving this: global optimization, geometric methods, physics-based ones and others. We want to introduce yet another smoothing approach by treating each geometric element as a player in a game: a quest for the best element quality. In other words, each player has the goal of becoming as regular as possible. The set of strategies for each element is given by all translations of its vertices. Ideally, he would like to quantify this regularity using a quality measure which corresponds to the utility function in game theory. Each player is aware of the other players' utility functions as well as their set of strategies, which is analogous to his own utility function and strategies. In the simplest case, the utility functions only depend on the regularity. In more complicated cases 
this utility function depends on the element size, the curvature, or even the solution to a differential equation.
Each one of them is trying to find a location, that makes the adjacent 
geometric elements more regular.
It is not our objective to develop a new smoothing method, which outperforms other smoothing methods, but to present a new interpretation of smoothing, which can address certain issues in a new framework, give them a different spin and find applications based on this approach. In particular, we want to see how equilibria and optimal constellations relate to each other: Could equilibria in a game behave better in some ways, even though do not necessarily optimize an objective function?


\section{Survey on mesh smoothing methods}

Meshes are built from geometric elements like tetrahedra, triangles, hexahedra, prisms, among others. They collectively represent objects like auto parts and are used for physical simulations. In order to allow for efficient and effective FEM simulations, all geometric elements need to be of good quality. The process of increasing the quality of all geometric elements is called mesh smoothing. Determining the quality with respect to FEM is mathematically challenging \cite{Shewchuk2002} and is based on strong bounds on interpolation error, matrix condition numbers, discretization error, and other application needs. A generic smoothing method therefore often ignores these issues and assumes that the degree of regularity is a simple and effective way of measuring the quality for geometric elements. In these cases, an ad hoc quality function is defined representing how close the geometric element is to being regular.

Mesh smoothing methods can be classified \cite{MeiTipper2013_SmoothingPlanarMeshes,Owen1998,Wilson2011_HexahedralSmoothing} as geometry-based \cite{Field1988,VartziotisWipperSchwald2009}, optimization-based \cite{Branets2005, FreitagJonesPlassman1995,Parthasarathy1991,Shivanna2010,LengZhangXu_NovelGeometricFlowDrivenApproach2012,SastryShontz2012_MeshQualityImprovement,BrewerDiachinKnuppLeurentMelander2003}, phyics-based \cite{Shimada2000_Bubbles} and combinations thereof \cite{CanannTristanoStaten1998,Freitag1997,ChenTristanoKwok2003}. In order to be more effective, these methods should be combined with topological modifications \cite{BossenHeckbert1996,FreitagOllivierGooch1997,KlingnerShewchuk2007}. Optimization-based methods often lend themselves to untangling algorithms \cite{Knupp2001EWC,Li2000,FreitagPlassmann2000,AmentaBernEppstein1999} and can be further differentiated as local or global optimization-based. Geometry-based methods like the Laplacian \cite{Field1988} and GETMe \cite{
VartziotisWipperSchwald2009} smoothings have the advantage of being fast, but have always been heuristic until we proved in \cite{VartziotisHimpel2014}, that a minor variation of the GETMe method presented in \cite{VartziotisWipperSchwald2009} generalizes to a local optimization-based mesh smoothing and untangling method for mixed-volume meshes.

In the present work we introduce a novel approach to mesh smoothing by using algorithmic game theory to model the mesh smoothing behavior of the individual geometric elements. Rather than optimizing a global mesh quality function we claim that a game theoretic equilibrium generally gives smoothing results, which are better suited for FEM. The underlying reason is that interdependencies of the geometric elements with respect to FEM cannot easily be captured by a single mesh quality function, and it is precisely these relationships between the elements which can be modeled by a game

\section{Game theory for geometric elements}

A game consists of a number of players, a set of strategies and a payoff function that associates a payoff to each player with a choice of strategies for each player. More formally, the game is a triple of
\begin{enumerate}
 \item a set $P $ of players $e_1, \ldots, e_n$,
 \item a set $S = S_1 \times \ldots \times S_n$ of strategy profiles consisting of ordered $n$-tuples of strategies $s_i \in S_i$ for each player $e_i$, $i=1,\dots, n$, and
 \item a pay-off function $u = (u_1,\ldots, u_n) \co S \to \R^n$ consisting of a utility function $u_i\co S \to \R$ for each player $e_i$.
\end{enumerate}

In mesh smoothing, each smoothing step is the result of a single game. The whole smoothing process is the result of an iterated game. Even though the players and strategies of a game might not change, the smoothing result and therefore the pay-off functions will change. It might be useful to formulate a game, in which the triple is allowed to change.
In the following sections we specify the game for geometric elements.

\subsection{The players}

\begin{figure}[ht]
\centering
\def\svgwidth{5cm}
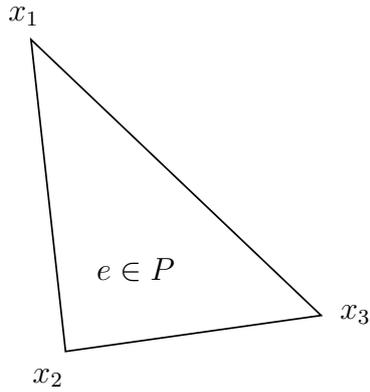
\caption{A typical player in the game\label{fig:player}} 
\end{figure}

Let us consider a mesh consisting of $n$ geometric elements $E$. These can for example be triangles, other polygons, tetrahedra, hexahedra or other polyhedra. The set of players is therefore given by $P:=E$. Let us assume that the combinatorial structure in the mesh given by the edges is fixed, so that $P$ does not change. Then each geometric element $e$ with $m_e$ vertices is determined by an $m_e$-tuple of coordinates $e = (x_e) = (x_1,\ldots,x_{m_e}) \in (\R^3)^{m_e}$. An example for a triangular player is given in Figure \ref{fig:player}.

\subsection{The strategy set}\label{sec:element_profiles}

\begin{figure}[ht]
\centering
\def\svgwidth{10cm}
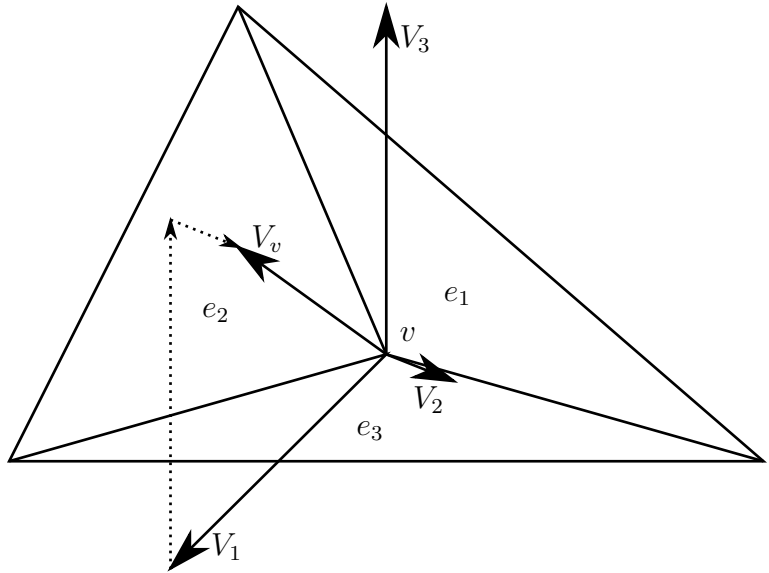
\caption{The tug-of-war by the mesh elements\label{fig:tugofwar}}
\end{figure}

The set of possible strategies $S_i$ for each player $e_i$ consists of all different ways (or a subset thereof) to push his vertices with some force or velocity. However, the vertices of adjacent geometric elements are identified, therefore the forces are coupled and cause the players' actions to affect each other like force vectors in a tug-of-war game. If $l$ players $e_i$ want to apply different force vectors $V_i \in S_i$ to a common vertex $v$, the resulting force vector applied to $v$ is given by $V_v = \sum V_i$. An example of a strategy profile $(V_1,V_2,V_3)$ with three players is shown in Figure \ref{fig:tugofwar}. In order for the players to decide, which strategy they should follow, they need to consider each others' utility functions. Given a choice of strategies for the other players, each player can make himself regular. This would result in a nonsense game, unless we modify the rules of the game. The simplest way to deal with this problem is to limit the set of strategies for each player to a 
finite number.

\subsection{The pay-off functions}

The utility function for each player is given by the element quality function which measures regularity, i.e. $u_i := q(e_i')$, where $e_i'$ is the result of applying the average force vector $V_v$ to each vertex $v$ of $e_i$ as shown in Figure \ref{fig:tugofwar}. Even though the game can be viewed as a multidirectional tug-of-war game, it is not, however, a multiplayer zero-sum game, because the overall pay-off is not constant for different strategies. To make matters worse, each player could make himself regular for any given choice of strategies of the other players, if we did not restrict the strategy profiles. This would result in a nonsensical game.

\subsection{Equilibrium}

We want to find equilibria in our game. If the players can use arbitrary force to move a vertex, there will be no equilibrium: Any player can modify his strategy to increase the pay-off for any choice of strategies for other players. This problem has been solved by considering only a finite set of strategies.

\section{GETMe Smoothing as a game}

GETMe smoothing has been developed and analyzed in a series of papers \cite{VartziotisAthanasiadisGoudasWipper2008,VartziotisWipperGETMeMixed2009,VartziotisHimpel2014,VartziotisBohnet2014}. GETMe smoothing can be considered not only by a spring model from physics perspective as in \cite{VartziotisBohnet2014b} but also from a game theory point of view. This might explain, why GETMe smoothing can give better results for FE simulations than global optimization based methods \cite{VartziotisWipperPapadrakakis2013}. It is the goal of the engineer to transform the mesh so that each element behaves well enough numerically with respect to FEM. In particular it is important for FEM, that all of the elements behave similarly well. This corresponds to an equilibrium with respect to FEM. To simplify this notion, algebraic quality measures are often used to determine how good a geometric element is. The numerical method is usually ignored. In global optimization-based methods the equilibrium aspect with respect to FEM is also ignored, because all local quality measures are simply combined to a (global) mesh quality function. There are several different ways of combining the local measures to a global one. For GETMe smoothing there is no global measure which guides GETMe. In the past implementations of GETMe however, the average of the mean ratio element qualities has been used as a criterion for ending the smoothing. This implies, that we can always make the result better by applying global optimization with respect to the average of the mean ratio qualities.

Based on the discussion above it would be better for FEM to let GETMe end, when it is close to an equilibrium rather than when some global mesh quality measure is maximal.

\section{Introducing GETMe Strategy Profiles}

A finite set of shifts given by geometric element transformations provides a particular set of strategy profiles. More specifically, if $T$ is a specific choice of GETMe transformation and $k\in \N$, we consider
\begin{enumerate}
 \item the set $P$ of geometric elements $e_1,\ldots, e_n \in E$,
 \item the set $S = S_1 \times \ldots \times S_n$ of strategy profiles, where $S_i = \{T^j\}_{j = 1\ldots, k}$, and
 \item the pay-off function $u = (u_1,\ldots, u_n)\co S \to \R^n$, where $u_i$ is an element quality measure for the result of $e_i$ after applying the transformation corresponding to the chosen strategy profile.
\end{enumerate}
This smoothing game needs to be repeated in order to optimize a mesh, because it represents only a single smoothing iteration.

\begin{figure}[ht]
\centering
\def\svgwidth{\textwidth/3-1.5mm}
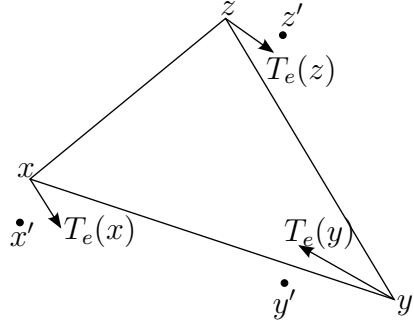
\caption{A GETMe transformation $T_e$ applied to a triangle $e$ computes ideal new coordinates $T_e(e)$ and moves $x$ to the average $x'=\frac{1}{|E_x|} \sum_{f \in E_x} T_f(x)$ over all triangles $f\in E$  containing $x$.\label{fig_getme}}
\end{figure}

Consider a triangle mesh embedded in three-dimensional Euclidean space. Let $E$ be the set of geometric elements in a mesh given as triples $e = (x,y,z) \in (\R^3)^3$ and $V$ the set of all vertices, that can be relocated on top of the geometric realization $M \subset \R^3$ of the mesh. Let $E_x$ be the subset of triangles attached to $x$. We view the elements $e \in E$ as players of a smoothing game. The strategies $S_i$ are interdependent. We might have the strategy to move a vertex $x\in e$ to a new position $T_e(x)$, but the actual coordinates are given by the average
\[x'=\frac{1}{|E_x|} \sum_{f \in E_x} T_f(x)\]
of all transformation results $T_f(x)$ for a choice of $T_f \in \cT_f$ associated to $f \in E_x$ as shown in Figure \ref{fig_getme}. Even though the strategy for $e$ is to move to $T_e(e)$, the new element will actually be $e'=(x',y',z')$.

\section{Preliminary numerical results}

Consider a simple planar mesh consisting of $n$ triangles, whose boundary is an $n$-gon. Every triangle can choose to apply the GETMe transformation $T$ $j$ times, $j = 0, \ldots, k$. The new coordinates are written as $x_j^{(s_j)} = T^{s_j} x_j$. The set of strategy profiles is $S = \prod_{i=1}^n S_i$, where $S_i = \{T^j\}_{j=1,\ldots,k}$. Let us identify a strategy profile $(T^{j_1},\ldots, T^{j_k}) \in S$ with the $n$-tuple $(j_1,\ldots, j_k) \subset \{0,\ldots,k\}^n$. The new coordinates $x_i^{(s)}$ are computed using the arithmetic mean
\[
x_i^{(s)}= \frac{1}{|N_i|}\sum_{j \in N_i} x_{i}^{(s_{j})}, 
\]
where $N_i$ is the index set of the triangles $e_j$ containing $x_i$. The utility function $u_i$ for each triangle $e_i=(x_{i_0},x_{i_1},x_{i_2})$ is given by $u_i(\sigma)=\frac{\min_{j,k \in \left\{0,1,2\right\}} \left\|x_{i_j}^{(\sigma)}-x_{i_k}^{(\sigma)}\right\|}{\max_{j,k \in \left\{0,1,2\right\}} \left\|x_{i_j}^{(\sigma)}-x_{i_k}^{(\sigma)}\right\|}$. A \emph{Nash equilibrium} $(s_i^*)_{i=1,\dots,n} \in \Pi_{i=1}^{n} S_n$ is the strategy, so t the value of $u_i$ is maximal, if the other triangles are not changing their strategy.

\subsection{Example 1: a simple mesh with 5 triangles}

\begin{figure}[h]
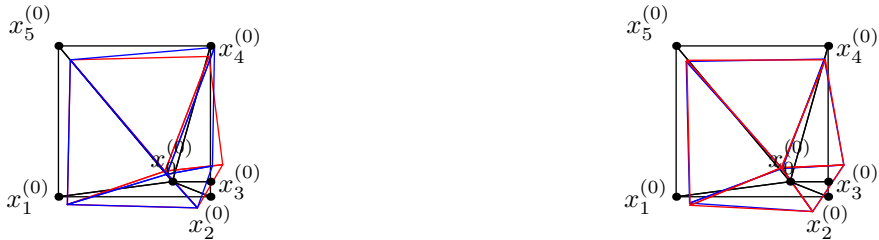

\begin{center}
\begin{minipage}{0.47\textwidth}
\centering
\includegraphics{beispiel1}
\end{minipage}
\begin{minipage}{0.47\textwidth}
\centering
\includegraphics{beispiel1b}
\end{minipage}
\caption{Example 1: initial mesh black, Nash equilibrium blue, best mesh quality red; on the left for $S = \{0,1,2\}^5$, on the right for $S = \{0,1,2,3\}^5$.\label{fig:beispiel1}}
\end{center} 
\end{figure}

The first example is a mesh with $5$ triangles (Figure~\ref{fig:beispiel1}). The best mesh quality 
$q_{\text{best}}=\frac{1}{5}\sum_{i=1}^{5} u_i(\sigma_{\text{best}})=0.651436$ is reached with the strategy profile $s_{\text{best}} = (2,2,1,2,2)$. The Nash equilibrium is $s^* = (2,2,2,2,2)$ and yields the mesh quality $q_{\text{Nash}}=0.648214$. Figure~\ref{fig:beispiel1} shows both results as well as the initial mesh. If we reduce the strategy profiles to $S = \left\{0,1\right\}^5$ the nash equilibrium also yields the maximal mesh quality. If we extend them to $S = \{0,1,2,3\}^5$, the nash equlibrium is $s^*= \{3,2,3,2,3\}$ with mesh quality $q_{\text{Nash}}=0.663681$, while the maximal mesh quality is $q_{\text{best}}=0.666673$ for the strategy profile $s_{\text{best}}=(3 2 3 3 2)$.

\subsection{Example 2: a perturbation of Example 1}

\begin{figure}[h]
\begin{center}
\includegraphics{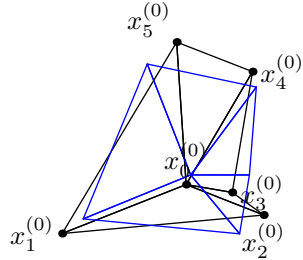}
\caption{Example 2: initial mesh black, nash equilibrium blue, $S = \{0,1,2\}^5$.\label{fig:beispiel2}}
\end{center}
\end{figure}

If we perturb Example 1, then the nash equilibrium has the maximal possible quality for the set of strategy profiles $S = \{0,1,2\}^5$ (see Figure~\ref{fig:beispiel2}. If we extend the set to $S = \{0,1,2,3\}^5$, the Nash and maximal quality strategy profiles are different: The Nash equilibrium is $s^* = (3,3,3,2,3)$, the best mesh quality is obtained via $s_{\text{best}}=(3, 3, 3, 2, 2,)$. The Difference between $q_{\text{Nash}}=0.680905$ and $q_{\text{best}}=0.687053$ is minimal with a relative difference $< 0.01$.

\subsection{Example 3: a simple mesh with 6 triangles}

\begin{figure}[h]
\begin{center}
\includegraphics{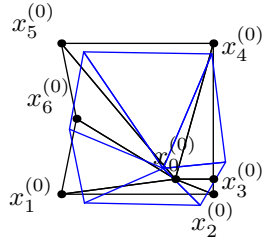}
\caption{Example 3: initial mesh black, Nash equilibrium blue, $S = \{0,1,2\}^6$.\label{fig:beispiel3}}
\end{center}
\end{figure}

For a simple mesh with 6 triangles as in Figure~\ref{fig:beispiel3} the Nash equilibrium attains maximal mesh quality with $s^* = s_{\text{best}}=(2 2 2 2 2 2)$. For $S = \{0,1,2,3\}^6$ with get $\sigma^*=\sigma_{\text{best}}=(3 2 3 2 3 3)$.

\subsection{Example 4: a simple mesh with 4 triangles}

In this example of a simple mesh with 4 polygons the initial quality is $q=0.423394$. With $S = \{0,1,2\}^4$ the Nash equilibrium $s^*=(0, 2, 2, 2)$ attains the maximal mesh quality $q_{\text{Nash}}=q_{\text{best}}=0.591098$. For $S =\{0,1,2,3\}^4$, however, we have $s^*=(3, 3, 3, 3)$ and $q_{\text{Nash}}=0.529091$, while we have $q_{\text{best}}=0.622306$ for $s_{\text{best}}=(0, 3, 3, 3)$. The difference between the Nash equilibrium and the best possible mesh is even more evident for $S=\{0,1,2,3,4\}^4$. We have $s^*=(4, 3, 3, 3)$ with $q_{\text{Nash}}=0.538821$ compared to $q_{\text{best}}=0.624241$ for $s_{\text{best}}=(0, 3, 4, 4)$. For $S =\{0,1,2,3,4,5\}^4$ as well as $S =\{0,1,2,3,4,5,6\}^4$ we have $s^*=(5 3 3 3)$, while the maximum remains at $s_{\text{best}}=(0, 3, 4, 4)$.

\begin{figure}[h]
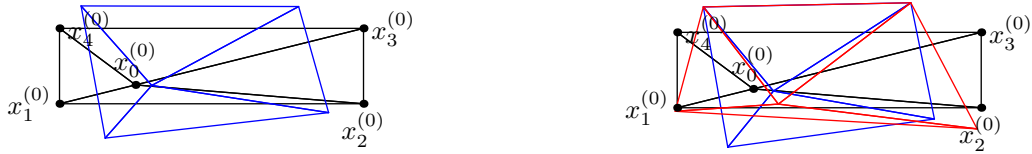

\begin{center}
\begin{minipage}{0.47\textwidth}
\includegraphics{beispiel4}
\end{minipage}
\begin{minipage}{0.47\textwidth}
\includegraphics{beispiel4b}
\end{minipage}
\caption{Example 4: initial mesh black, Nash equilibrium blue, best quality mesh red, $S = \{0,1,2\}^4$ on the left $S = \{0,1,2,3\}^4$ on the right.\label{fig:beispiel4}}
\end{center} 
\end{figure}  

\begin{figure}[h]
\begin{center}
\includegraphics[width=0.8\textwidth]{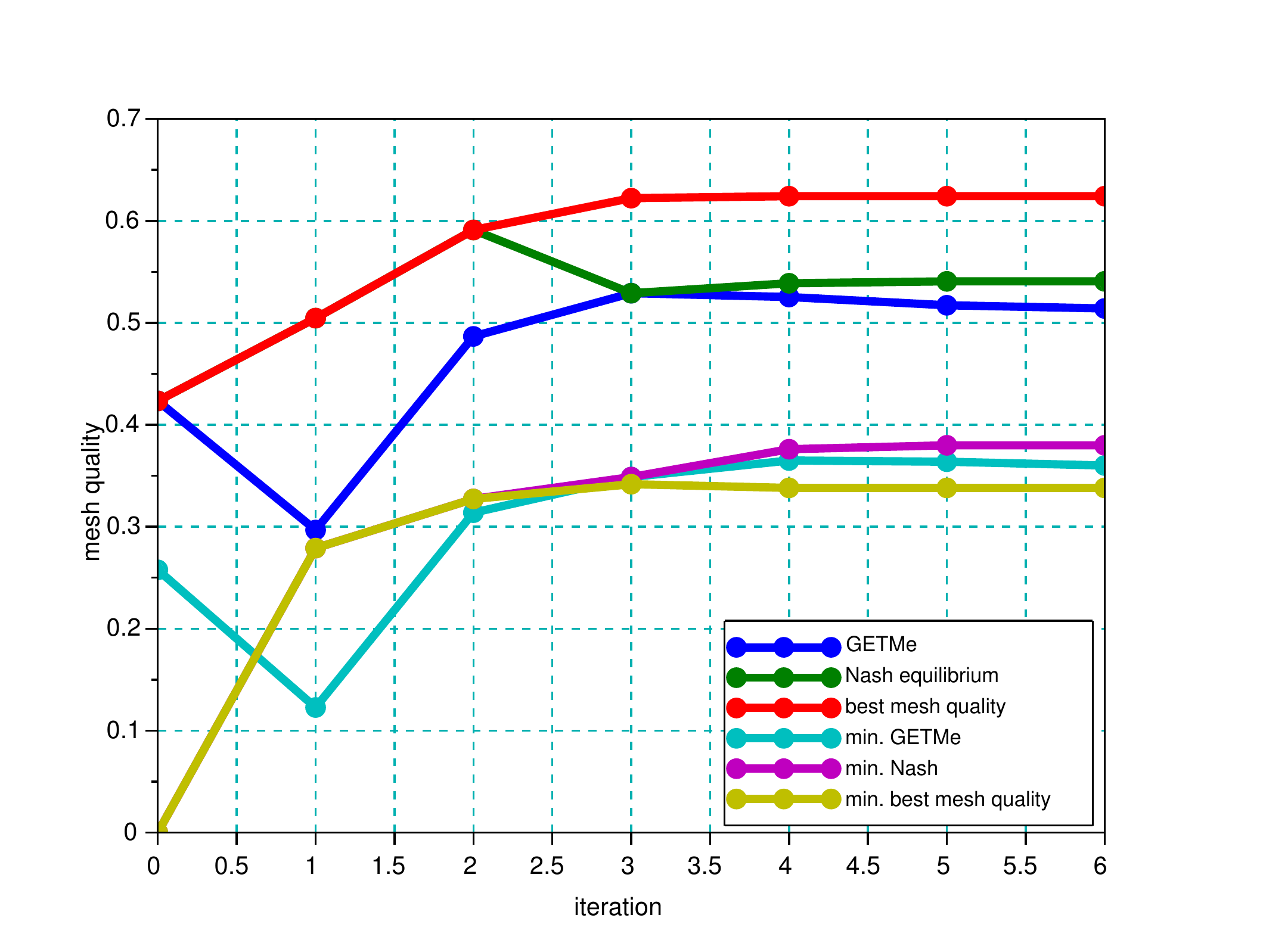}
\caption{Vergleich über die Entwicklung der Netzqualität in Abhängigkeit der Iterationen\label{fig:auswertung}}
\end{center} 
\end{figure}

If we compare the strategy of the Nash equilibrium with the usual mesh smoothing strategy $s_{\text{GETMe}}=(k, k, k, k)$ for $S=\{0,\dots,k\}^4$ (see Figure~\ref{fig:auswertung}) we can see that the Nash equilibrium yields a higher mash quality {\bf Usual mesh smoothing strategy?}. Furthermore, the mesh quality reduces for $k >4$, while the mesh quality of the Nash equilibrium increases and eventually stays constant. Figure~\ref{fig:beispiel4} shows that the mesh with the best mesh quality does not necessarily look best: The optimal mesh quality has been reached at the cost of a very bad triangle. The quality of such a mesh therefore has a relatively small minimal quality. 

\subsection{A more complex mesh}

The smoothing game falls in the category of graphical games. These are games, in which the utilities of each player depends on the actions of very few other players. In our case, each players utility only depends on his neighbors. Unlike normal form games, the problem of finding a pure Nash equilibrium in graphical games (if one exists) is NP-complete \cite{GottlobGrecoScarcello2005}. The problem of finding a (possibly mixed) Nash equilibrium in a graphical game is PPAD-complete \cite{DaskalakisFabrikantPapadimitriou2006}. Finding a correlated equilibrium of a graphical game can be done in polynomial time, and for a graph with a bounded treewidth, this is also true for finding an optimal correlated equilibrium \cite{PapadimitriouRoughgarden2008}.

Instead of trying to implement an existing algorithm from algorithmic game theory to find a (possibly mixed) Nash equilibrium we are content with trying to efficiently finding a Nash equilibrium, stopping when we have found one or aborting, when it seems appropriate. It will be interesting to compare the result of this approach with other smoothing algorithms. We expect Algorithm \ref{alg:gamesmooth} this approach based on GETMe strategies to give a result which looks similar to the result of the usual GETMe smoothing.

\begin{algorithm}
\caption{Game Smoothing Algorithm}\label{alg:gamesmooth}
\begin{algorithmic}[1]
\Procedure{Smooth mesh}{q}
\State $\textit{ElementListByQuality} \gets \text{sort } \textit{ElementList} \text{ by its quality}$
\State $\textit{element} \gets \textit{ElementListByQuality}$
\While {$\text{quality}(\textit{element})<q$}
\State {\it 1-level neighborhood} $\gets$ {\it element} and its neighbors (fixed boundary)
\State Find Nash equilibrium({\it 1-level neighborhood with fixed boundary}).
\State Save the new coordinates
\State Adjust the list positions in {\it ElementListByQuality} 
\State $\textit{element} \gets \textit{ElementListByQuality}$
\EndWhile
\State Make choice permanent
\EndProcedure
\end{algorithmic}
\end{algorithm}

\section{Conclusion}
\begin{enumerate}
\item The Nash equilibrium for the set of strategy profiles $S = \{0,\ldots, n\}^n$ generally does not correspond to the strategy $(n,n,\ldots,n)$, which is usually used for geometric mesh smoothing methods.
\item The Nash equilibrium generally does not give the best possible mesh quality.
\item The minimal element quality seems to be better for the Nash equilibrium compared to the best quality mesh. The constellations with maximal quality therefore look worse than the Nash equilibra.
\end{enumerate}

\bibliographystyle{alpha}
\bibliography{literature}

\newcommand{\etalchar}[1]{$^{#1}$}
\begin{thebibliography}{VAGW08}

\bibitem[ABE99]{AmentaBernEppstein1999}
Nina Amenta, Marshall Bern, and David Eppstein.
\newblock {Optimal Point Placement for Mesh Smoothing}.
\newblock {\em Journal of Algorithms}, 30:302--322, 1999.

\bibitem[BDK{\etalchar{+}}03]{BrewerDiachinKnuppLeurentMelander2003}
Michael Brewer, Lori A.~Freitag Diachin, Patrick~M. Knupp, Thomas Leurent, and
  Darryl Melander.
\newblock The {M}esquite mesh quality improvement toolkit.
\newblock In {\em Proceedings of the 12th International Meshing Roundtable},
  pages 239--250, 2003.

\bibitem[BH96]{BossenHeckbert1996}
Frank~J. Bossen and Paul~S. Heckbert.
\newblock {A Pliant Method for Anisotropic Mesh Generation}.
\newblock In {\em Proceedings of the 5th International Meshing Roundtable},
  1996.

\bibitem[Bra05]{Branets2005}
Larisa~Vladimirovna Branets.
\newblock {\em A variational grid optimization method based on a local cell
  quality metric}.
\newblock PhD thesis, Austin, TX, USA, 2005.
\newblock AAI3187661.

\bibitem[CTK03]{ChenTristanoKwok2003}
Zhijian Chen, Joseph~R. Tristano, and Wa~Kwok.
\newblock {Combined Laplacian and Optimization-based Smoothing for Quadratic
  Mixed Surface Meshes}.
\newblock In {\em Proceedings of the 12th International Meshing Roundtable},
  2003.

\bibitem[CTS98]{CanannTristanoStaten1998}
Scott~A. Canann, Joseph~R. Tristano, and Matthew~L. Staten.
\newblock An approach to combined laplacian and optimization-based smoothing
  for triangular, quadrilateral, and quad-dominant meshes.
\newblock In {\em Proceedings of the 7th International Meshing Roundtable},
  pages 479--494, 1998.

\bibitem[DFP06]{DaskalakisFabrikantPapadimitriou2006}
Constantinos Daskalakis, Alex Fabrikant, and Christos~H. Papadimitriou.
\newblock The game world is flat: the complexity of {N}ash equilibria in
  succinct games.
\newblock In {\em Automata, languages and programming. {P}art {I}}, volume 4051
  of {\em Lecture Notes in Comput. Sci.}, pages 513--524. Springer, Berlin,
  2006.

\bibitem[Fie88]{Field1988}
David~A. Field.
\newblock {Laplacian smoothing and Delaunay triangulations}.
\newblock {\em Communications in Applied Numerical Methods}, 4(6):709--712,
  1988.

\bibitem[FJP95]{FreitagJonesPlassman1995}
Lori~A. Freitag, Mark Jones, and Paul Plassmann.
\newblock {An Efficient Parallel Algorithm for Mesh Smoothing}.
\newblock In {\em Proceedings of the 4th International Meshing Roundtable},
  pages 47--58, 1995.

\bibitem[FOG97]{FreitagOllivierGooch1997}
Lori~A. Freitag and Carl Ollivier-Gooch.
\newblock {T}etrahedral {M}esh {I}mprovement {U}sing {S}wapping and
  {S}moothing.
\newblock {\em International Journal for Numerical Methods in Engineering},
  40(21):3979--4002, 1997.

\bibitem[FP00]{FreitagPlassmann2000}
Lori~A. Freitag and Paul Plassmann.
\newblock Local optimization-based simplicial mesh untangling and improvement.
\newblock {\em International Journal of Numerical Methods in Engineering},
  49(1--2):109--125, 2000.

\bibitem[Fre97]{Freitag1997}
Lori~A. Freitag.
\newblock {On combining Laplacian and optimization-based mesh smoothing
  techniques}.
\newblock In {\em Trends in Unstructured Mesh Generation}, pages 37--43, 1997.

\bibitem[GGS05]{GottlobGrecoScarcello2005}
Georg Gottlob, Gianluigi Greco, and Francesco Scarcello.
\newblock Pure {N}ash equilibria: hard and easy games.
\newblock {\em J. Artificial Intelligence Res.}, 24:357--406 (electronic),
  2005.

\bibitem[Knu01]{Knupp2001EWC}
Patrick~M. Knupp.
\newblock {H}exahedral and {T}etrahedral {M}esh {U}ntangling.
\newblock {\em Engineering with Computers}, 17(3):261--268, 2001.

\bibitem[KS07]{KlingnerShewchuk2007}
Bryan~Matthew Klingner and Jonathan~Richard Shewchuk.
\newblock {Aggressive Tetrahedral Mesh Improvement}.
\newblock In {\em Proceedings of the 16th International Meshing Roundtable},
  pages 3--23, 2007.

\bibitem[LWH{\etalchar{+}}00]{Li2000}
T.S. Li, S.M. Wong, Y.C. Hon, C.G. Armstrong, and R.M. McKeag.
\newblock Smoothing by optimisation for a quadrilateral mesh with invalid
  elements.
\newblock {\em Finite Elements in Analysis and Design}, 34(1):37 -- 60, 2000.

\bibitem[LXZQ12]{LengZhangXu_NovelGeometricFlowDrivenApproach2012}
Juelin Leng, Guoliang Xu, Yongjie Zhang, and Jin Qian.
\newblock A {N}ovel {G}eometric {F}low-{D}riven {A}pproach for {Q}uality
  {I}mprovement of {S}egmented {T}etrahedral {M}eshes.
\newblock In William~Roshan Quadros, editor, {\em Proceedings of the 20th
  International Meshing Roundtable}, pages 347--364. Springer Publishing
  Company, Incorporated, 2012.

\bibitem[MTX13]{MeiTipper2013_SmoothingPlanarMeshes}
Gang Mei, John~C. Tipper, and Nengxiong Xu.
\newblock The {M}odified {D}irect {M}ethod: An {I}terative {A}pproach for
  {S}moothing {P}lanar {M}eshes.
\newblock In {\em ICCS}, pages 2436--2439, 2013.

\bibitem[Owe98]{Owen1998}
Steven~J. Owen.
\newblock A {S}urvey of {U}nstructured {M}esh {G}eneration {T}echnology.
\newblock In {\em Proceedings of the 7th International Meshing Roundtable},
  pages 239--267, 1998.

\bibitem[PK91]{Parthasarathy1991}
V.N. Parthasarathy and Srinivas Kodiyalam.
\newblock A constrained optimization approach to finite element mesh smoothing.
\newblock {\em Finite Elements in Analysis and Design}, 9(4):309 -- 320, 1991.

\bibitem[PR08]{PapadimitriouRoughgarden2008}
Christos~H. Papadimitriou and Tim Roughgarden.
\newblock Computing correlated equilibria in multi-player games.
\newblock {\em J. ACM}, 55(3):Art. 14, 29, 2008.

\bibitem[SGM10]{Shivanna2010}
Kiran Shivanna, Nicole Grosland, and Vincent Magnotta.
\newblock An {A}nalytical {F}ramework for {Q}uadrilateral {S}urface {M}esh
  {I}mprovement with an {U}nderlying {T}riangulated {S}urface {D}efinition.
\newblock In Suzanne Shontz, editor, {\em Proceedings of the 19th International
  Meshing Roundtable}, pages 85--102. Springer Berlin Heidelberg, 2010.

\bibitem[She02]{Shewchuk2002}
Jonathan~R. Shewchuk.
\newblock {What is a Good Linear Element? Interpolation, Conditioning, and
  Quality Measures}.
\newblock In {\em Proceedings of the 11th International Meshing Roundtable},
  pages 115--126, 2002.

\bibitem[SS12]{SastryShontz2012_MeshQualityImprovement}
Shankar~Prasad Sastry and Suzanne~M. Shontz.
\newblock Performance characterization of nonlinear optimization methods for
  mesh quality improvement.
\newblock {\em Eng. with Comput.}, 28(3):269--286, July 2012.

\bibitem[SYI00]{Shimada2000_Bubbles}
Kenji Shimada, Atsushi Yamada, and Takayuki Itoh.
\newblock Anisotropic {T}riangulation of {P}arametric {S}urfaces via {C}lose
  {P}acking of {E}llipsoids.
\newblock {\em Internat. J. Comput. Geom. Appl.}, 10(4):417--440, 2000.
\newblock Selected papers from the Sixth International Meshing Roundtable, Part
  II (Park City, UT, 1997).

\bibitem[VAGW08]{VartziotisAthanasiadisGoudasWipper2008}
Dimitris Vartziotis, Theodoros Athanasiadis, Iraklis Goudas, and Joachim
  Wipper.
\newblock {Mesh smoothing using the Geometric Element Transformation Method}.
\newblock {\em Comput. Methods Appl. Mech. Engrg.}, 197(45--48):3760--3767,
  2008.

\bibitem[VB14a]{VartziotisBohnet2014}
Dimitris Vartziotis and Doris Bohnet.
\newblock Convergence properties of a geometric mesh smoothing algorithm.
\newblock {\em arXiv:1411.3869 [math.NA]}, 2014.

\bibitem[VB14b]{VartziotisBohnet2014b}
Dimitris Vartziotis and Doris Bohnet.
\newblock A geometric mesh smoothing algorithm related to damped oscillations.
\newblock {\em arXiv:1411.4390 [math.NA]}, 2014.

\bibitem[VH14]{VartziotisHimpel2014}
Dimitris Vartziotis and Benjamin Himpel.
\newblock Efficient mesh optimization using the gradient flow of the mean
  volume.
\newblock {\em SIAM Journal on Numerical Analysis}, 52(2):1050--1075, 2014.

\bibitem[VW09]{VartziotisWipperGETMeMixed2009}
Dimitris Vartziotis and Joachim Wipper.
\newblock {The Geometric Element Transformation Method for Mixed Mesh
  Smoothing}.
\newblock {\em Eng. Comput.}, 25(3):287--301, 2009.

\bibitem[VWP13]{VartziotisWipperPapadrakakis2013}
Dimitris Vartziotis, Joachim Wipper, and Manolis Papadrakakis.
\newblock Improving mesh quality and finite element solution accuracy by
  {GETM}e smoothing in solving the {P}oisson equation.
\newblock {\em Finite Elem. Anal. Des.}, 66:36--52, 2013.

\bibitem[VWS09]{VartziotisWipperSchwald2009}
Dimitris Vartziotis, Joachim Wipper, and Bernd Schwald.
\newblock The geometric element transformation method for tetrahedral mesh
  smoothing.
\newblock {\em Comput. Methods Appl. Mech. Engrg.}, 199(1-4):169--182, 2009.

\bibitem[Wil11]{Wilson2011_HexahedralSmoothing}
Thomas~James Wilson.
\newblock Simultaneous {U}ntangling and {S}moothing of {H}exahedral {M}eshes.
\newblock Master's thesis, Universitat Polit{\`e}cnica de Catalunya, Spain,
  2011.

\end{thebibliography}

\end{document}